\documentclass[letterpaper, 10 pt, conference]{ieeeconf}  
\IEEEoverridecommandlockouts
\usepackage{amsmath}
\usepackage{amsfonts}
\usepackage{graphicx}
\usepackage{xcolor}
\usepackage{figsize}
\usepackage{upgreek}
\usepackage{amssymb}
\usepackage{algorithmic}
\usepackage[linesnumbered,ruled,vlined]{algorithm2e}
\usepackage{siunitx}

\newtheorem{definition}{\bf Definition}
\newtheorem{theorem}{\bf Theorem}
\newtheorem{lemma}{\bf Lemma}
\newtheorem{assumption}{\bf Assumption}
\newtheorem{proposition}{\bf Proposition}
\newtheorem{remark}{\bf Remark}
\newtheorem{notation}{Notation}

\newcommand\norm[1]{\left\lVert#1\right\rVert}

\title{\LARGE \bf
RNN Controller for Lane-Keeping Systems with Robustness and Safety Verification}
\author{Ying Shuai Quan$^{1}$, Jin Sung Kim$^{1}$, and Chung Choo Chung$^{1^\dag}$
\thanks{*This work was supported by the National Research Foundation of Korea(NRF) grant funded by the Korea government(MSIT) (No. 2021R1A2C2009908, Data-Driven Optimized Autonomous Driving Technology Using Open Set Classification Method).}
\thanks{$^{1}$Y. S. Quan, J. S. Kim, and C. C. Chung are with Dept. of Electrical Engineering, Hanyang University, Seoul 04763, Republic of Korea.
        {\tt\small {\{ysquan, jskim06,  cchung\}@hanyang.ac.kr}}}
}
\begin{document}
\maketitle
\thispagestyle{empty}
\pagestyle{empty}
\begin{abstract}
This paper proposes a Recurrent Neural Network (RNN) controller for lane-keeping systems, effectively handling model uncertainties and disturbances. First, quadratic constraints cover the nonlinearities brought by the RNN controller, and the linear fractional transformation method models the dynamics of system uncertainties. Second, we prove the robust stability of the lane-keeping system in the presence of uncertain vehicle speed using a linear matrix inequality. Then, we define a reachable set for the lane-keeping system. Finally, to confirm the safety of the lane-keeping system with tracking error bound, we formulate semidefinite programming to approximate the outer set of the reachable set. Numerical experiments demonstrate that this
approach confirms the stabilizing RNN controller and validates the safety with an untrained dataset with untrained varying road curvatures.
\end{abstract}
%
\section{INTRODUCTION}
In general, it is difficult for Neural Network (NN)-based controllers to provide  guaranteed stability due to the complex structures with nonlinearites~\cite{fazlyab2020safety,yin2021stability}.
In the recent literature, classical robust control methods-based analysis and synthesis of NNs have been actively discussed, where Quadratic Constraint (QC) multipliers are employed to over-approximate the dynamics of nonlinear activation functions of NNs~\cite{fazlyab2020safety,yin2021stability, revay2020convex, gu2022recurrent, junnarkar2022synthesis}.
The QC-based methods enable the robustness analysis of the NN-controlled system through Linear Matrix Inequality (LMI) optimization.
Bounds on the Lipschitz constant of feedforward NNs and Recurrent Neural Networks (RNNs) are guaranteed in~\cite{fazlyab2020safety,revay2020convex}.
~\cite{yin2021stability,gu2022recurrent, junnarkar2022synthesis} provide robust stability conditions for partially observed systems stabilized by NN and RNN controllers.

Studies have focused on handling perturbations such as noises and adversarial attacks~\cite{fazlyab2020safety}. However, there is still a lot of potential for exploring the incorporation of dynamic uncertainties, especially when using RNN as a controller~\cite{gu2022recurrent, junnarkar2022synthesis}.
To explicitly analyze the dynamic uncertainties, a commonly applied technique in classical robust control theory is the Linear Fractional Transformation (LFT) method, which provides a convenient means to capture the system uncertainty and represent it in the form of state-space systems~\cite{khalil1996robust}.
The state-space modeled uncertainties can then be integrated with LMI optimizations, enabling analyzing robust stability against various uncertainty classes in a straightforward way~\cite{veenman2016robust}.

Our focus in this paper is centered around the utilization of an RNN controller  for a lane-keeping system. During realistic driving situations, it is common for experienced human drivers to adjust the vehicle's longitudinal speed in response to changing road conditions, such as road curvatures. However, this speed variation can cause model uncertainty in the vehicle's lateral dynamics models. To address this issue, efficient and high-performing optimal control methods, such as Model Predictive Control (MPC), have been proposed for lane-keeping maneuvers~\cite{choi2021horizonwise}. However, accurately predicting future speed variation not available during the prediction horizon limits the application of Linear Time-Varying (LTV)-MPC.
The method proposed in~\cite{choi2021horizonwise} addresses this problem and resolves the issue under a particular assumption, i.e., uniform compensation condition, which limits its applicability.
In addition, the implementation of LTV-MPC as chassis control may be impeded by the long and even uncertain computation time. It was also shown that an RNN controller learned from multiple Linear Time-Invariant (LTI)-MPCs for lane changing may be a potential solution~\cite{quan2020multi}, in which robustness and stability under varying conditions, however, are not discussed.

In this context, to train an RNN controller, we utilize LTV-MPC to generate the training datasets, in which the speed variation within the prediction horizon is available.
The contributions of this paper are two-fold.
\begin{itemize}
	\item For the robustness analysis with the model uncertainty, we provide an analysis method of the RNN controller based on classical robust control theory.
Robust stability analysis for the RNN-controlled system is formulated into an LMI problem.
QC describes the nonlinearities brought by the RNN controller, and the LFT method is applied to cover the dynamics of system uncertainty.
The QC and LFT-generated state-space model are later combined into an LMI optimization to provide robustness analysis for the closed-loop system.
\item In addition, to ensure the safety of the RNN-controlled system when encountering untrained road conditions such as road curvatures, we provide an outer approximation of the reachable set for the system state, which enables online safety verification.
The reachable set is obtained by solving a Semidefinite Programming (SDP) problem to reduce the conservativeness.
\end{itemize}
The performance of the proposed RNN control method is validated with simulations of lane-keeping maneuvers even under a previously unknown speed variation at test highway roads with various curvatures.
\begin{notation}
$\mathbb{S}^n, \mathbb{S}^n_+$, $\mathbb{S}^n_{++}$ denote the sets of $n$-by-$n$ symmetric, positive semidefinite and positive definite matrices, respectively.
	$\mathbb{D}^n_+$, $\mathbb{D}^n_{++}$ denote the set of n-by-n diagonal positive semidefinite, and diagonal positive definite matrices.
	We denote the Euclidean norm as $\norm{x} = \sqrt{x^T x}$.
\end{notation}
\section{Preliminaries}
\begin{figure}[t]
    \centering
\includegraphics[width=0.95\columnwidth]{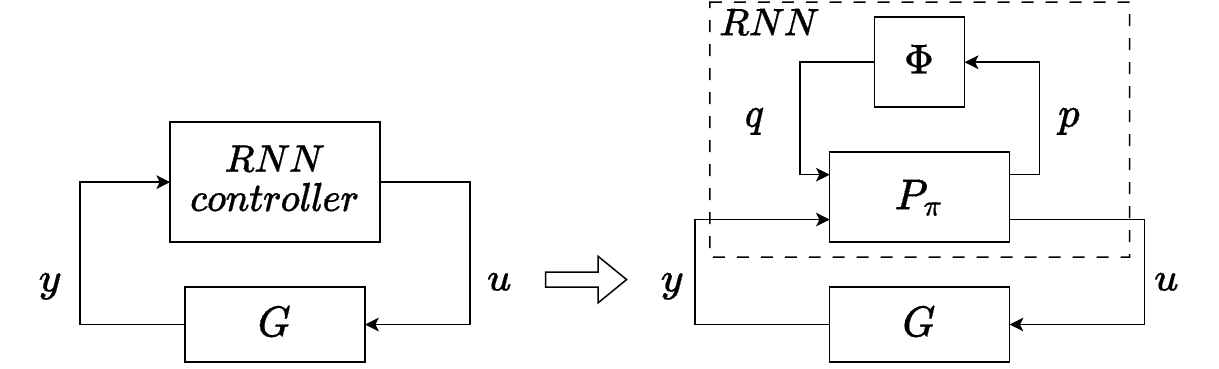}
		\caption{Feedback system consisting of plant $G$ and RNN controller. RNN controller is interpreted as an interconnection of $P_\pi$ and $\Phi$.}
\label{fig:fig1}
\end{figure}
\subsection{Problem formulation}
In this paper, we consider a feedback system consisting of a plant $G$ describing the vehicle lateral dynamics and an RNN controller stabilizing the system, as shown in Fig~\ref{fig:fig1}.
\subsubsection{Vehicle lateral dynamics}
We consider an empirical kinematic model for the vehicle lateral dynamics that describes lane-keeping errors at a look-ahead distance $L$, assuming no slip angles exist.
When performing lane-keeping maneuvers, a kinematic model is considered sufficient to describe vehicle lateral motion as tire slip and sideslip angles are small~\cite{kang2018multirate}.
Besides, using the kinematic model avoids involving uncertain vehicle parameters such as the tire-road conditions or the cornering stiffness~\cite{kang2018linear}.
A partially observed LTI system $G$ based on the empirical kinematic model in the discrete-time domain is given:
\begin{equation}
	\begin{split}
		G =
		\begin{cases}
			x_{k+1} &= A_Gx_{k}+B_G{}_1 u_{k}+B_G{}_2 \varphi_{k}\\
		    y_{k} &= C_G x_{k}
		\end{cases}
	\end{split}
	\label{eq:G}
\end{equation}
with
\begin{equation*}
	\begin{split}
		&x= {\begin{bmatrix} e_{yL} & \dot{e}_y & e_{\psi} & \dot{\psi} \end{bmatrix}}^T,
		u= \delta,~
		\varphi =
		\begin{bmatrix}
		\dot{\psi}^d\\
		e_{\psi L} - e_{\psi}
		\end{bmatrix},\\
		&A_G=
		\begin{bmatrix}
		  1 & T & 0 & -TL \\
		  0 & 1-\epsilon & \epsilon V_x & 0  \\
		  0 & 0 & 1 & -T \\
		  0 & 0 & 0 & 1-T/\tau_\psi \\
		\end{bmatrix},
		B_G{}_1 =
		\begin{bmatrix}
		  0 \\
		  -\epsilon\frac{l_r}{l_f + l_r}V_x\\
		  0 \\
		  \frac{T}{(l_f + l_r)\tau_\psi} V_x
		\end{bmatrix},\\
		&B_G{}_2 =
		\begin{bmatrix}
 		 TL & TV_x\\ 0 & 0 \\ T & 0 \\ 0 & 0 \\
		\end{bmatrix},
		C_G =
		\begin{bmatrix}
			1 & 0 & 0 & 0\\
			0 & 0 & 1 & 0
		\end{bmatrix},
	\end{split}
\end{equation*}
where
$x$ is the system state,
$u$ is the control input,
$\varphi$ is the external input caused by the winding road disturbance
and
$y$ is the observed state.
$e_{yL}$ and $e_{\psi L}$ is the lateral lane center offset and yaw angle error at the look-ahead point,
$\dot{e}_y$ is the time derivative of lateral position error with regard to reference,
$e_{\psi}$ is the yaw angle error with regard to the road,
$\dot{\psi}$ and $\dot{\psi}^d$ is the actual and desired yaw rate of the vehicle,
$\delta$ is the steering angle,
$V_x$ is the longitudinal velocity,
$l$ ($l_f$,$l_r$) is the longitudinal distance from the center of gravity to (front, rear) tires,
$T$ is the sampling time,
$0 \leq \epsilon \leq 1$ is a relaxation factor which describes the relationship between $\dot{e}_{\psi}$ and other factors, and $\tau_\psi$ is determined by experiments.
Notice that in this paper, we assume that the external input $\varphi$ can be estimated by the road lane curve~\cite{choi2021horizonwise}.
\subsubsection{RNN controller}
\begin{definition}
A function $\varphi: \mathbb{R}\rightarrow \mathbb{R}$ is sector-bounded such that $\varphi\in \text{sec}[\alpha, \beta]$ with $\alpha < \beta$ if  \begin{equation}
	(\varphi(p)-\alpha p)\cdot(\beta p-\varphi(p)) \geq 0, \forall p \in \mathbb{R}.
	\end{equation}
\end{definition}
The RNN controller can be interpreted as an interconnection of an LTI system $P_\pi$ and activation functions $\Phi: \mathbb{R}^{n_\Phi} \rightarrow \mathbb{R}^{n_\Phi}$, formulated as;
	\begin{equation}
	\begin{split}
		P_\pi &=
    \begin{cases}
      {\xi}_{k+1} &= {A}{\xi}_{k} + {B}_1 q_k + {B}_2 {y}_k \\
		{u}_{k} &= {C}_1{\xi}_{k} + {D}_{11} q_k + {D}_{12} {y}_k  \\
		{p}_{k} &= {C}_2{\xi}_{k} + {D}_{22} y_k
    \end{cases}\\
    q_k &= \Phi (p_k),
	\end{split}
	\end{equation}
where the combined nonlinearity $\Phi$ is element-wise:
\begin{equation}
	\Phi:= [\varphi_1(p_1),\cdots,\varphi_{n_\phi}(p_{n_\phi})]^T,
\end{equation}
with each $\varphi_i \in \text{sec}[\alpha_i, \beta_i]$  the $i$-th scalar activation function.
The activation function is assumed to have a fixed point at the origin, i.e., $\Phi(0) = 0$.
Assuming that each neuron $i$ contains a sector-bounded activation function $\varphi_i$, vectors $\alpha_\phi, \beta_\phi\in\mathbb{R}^{n_\phi}$ are stacked up with these sectors to establish the following QCs, where $\alpha_\phi = [\alpha_1,\cdots,\alpha_{n_\phi}]$ and $\beta_\phi = [\beta_1,\cdots,\beta_{n_\phi}]$.
\begin{lemma}
	Suppose that $\Phi$ satisfies the sector bound $[\alpha_\phi, \beta_\phi]$ element-wise with $\alpha_\phi, \beta_\phi\in\mathbb{R}^{n_\phi}$ and $\alpha_\phi \leq \beta_\phi$. For $\Lambda\in \mathbb{D}^{n_\phi}_{+}$ and for all $p \in\mathbb{R}^{n_\phi}$ and  $q=\Phi(p)$, it holds that
	 \begin{equation}
	\begin{bmatrix}
		p \\ q
	\end{bmatrix}^T
	\begin{bmatrix}
		-2A_\phi B_\phi \Lambda & (A_\phi+B_\phi)\Lambda \\
		(A_\phi+B_\phi)\Lambda & -2\Lambda,
	\end{bmatrix}
	\begin{bmatrix}
		p \\ q
	\end{bmatrix} \geq 0,
\end{equation}
where
$A_\phi = diag(\alpha_\phi), B_\phi = diag(\beta_\phi)$.
\end{lemma}
For the details of the proof, readers are referred to~\cite{fazlyab2020safety}.

To simplify the RNN analysis, we then perform a loop transformation that shifts and scales the activation function $\Phi$ to a new activation function $\tilde\Phi$ sector-bounded by $\text{sec}[-\textbf{1}_{n_\Phi\times 1}, \textbf{1}_{n_\Phi\times 1}]$~\cite{gu2022recurrent,junnarkar2022synthesis}, such that the following QC is satisfied:
\begin{equation}
	\begin{bmatrix}
		p \\ q
	\end{bmatrix}^T
	\begin{bmatrix}
		\Lambda & 0\\
		0 & -\Lambda
	\end{bmatrix}
	\begin{bmatrix}
		p \\ q
	\end{bmatrix} \geq 0.
	\label{eq:QC}
\end{equation}
\begin{figure}[t]
    \centering
\includegraphics[width=0.95\columnwidth]{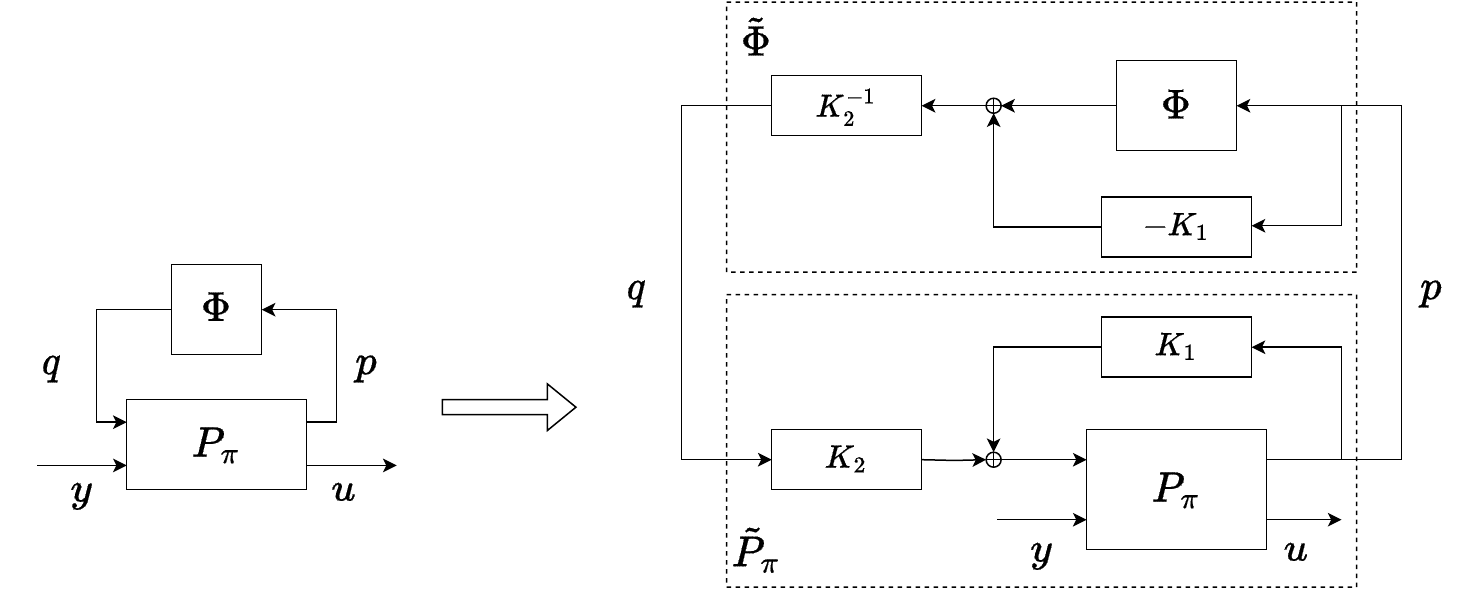}
		\caption{Loop transformation.}
		\label{fig:fig2}
\end{figure}
As is shown in Fig~\ref{fig:fig2}, the new representation of the RNN is then written as
	\begin{equation}
		\tilde{P}_\pi =
    \begin{cases}
      {\xi}_{k+1} &= \tilde{A}{\xi}_{k} + \tilde{B}_1 q_k + \tilde{B}_2 {y}_k \\
		{u}_{k} &= \tilde{C}_1{\xi}_{k} + \tilde{D}_{11} q_k + \tilde{D}_{12} {y}_k\\
		{p}_{k} &= {C}_2{\xi}_{k} + {D}_{22} y_k
    \end{cases}
	\end{equation}
	\begin{equation}
		\begin{split}
			q_k &= \tilde\Phi (p_k),
		\end{split}
	\end{equation}
	with
	\begin{equation*}
		\begin{split}&K_1:=\frac{A_\phi+B_\phi}{2},~
		    K_2 := \frac{B_\phi-A_\phi}{2},\\
		    &\tilde{A} = A+B_1K_1C_2,~
		    \tilde{B}_1 = B_1K_2,~
		    \tilde{B}_2 = B_2 + B_1K_1D_{22},\\
		    &\tilde{C}_1 = C_1+D_{11}K_1C_2,
		    \tilde{D}_{11} = D_{11}K_2,\\
		    &\tilde{D}_{12} = D_{12}+D_{11}K_1D_{22},~
		    \tilde\Phi = K^{-1}_2\Phi-K_1K^{-1}_2.
	\end{split}
	\end{equation*}	
Defining
$r:=
\begin{bmatrix}
p^T
&
q^T
\end{bmatrix}^T
$, the vector $r$ can be represented as
\begin{align}
\begin{split}
r = D_{rp}p + D_{rq}q,
\end{split}
\label{eq:define_r}
\end{align}
where
$D_{rp} =
\begin{bmatrix}
I_{n_\Phi} & 0
\end{bmatrix}^T
$, and
$D_{rq} =
\begin{bmatrix}
0 & I_{n_\Phi}
\end{bmatrix}^T
$.
$r$ will be later introduced into the LMI formulation for the robustness analysis.
\section{Robustness Analysis}
\subsection{Uncertain vehicle lateral dynamics}
In realistic driving situations, experienced human drivers tend to adjust the vehicle longitudinal speed to cope with different road conditions, such as road curvatures.
Therefore, we consider parameter uncertainty $\delta V_x$ in $V_x$ for the vehicle lateral dynamics.
Then (\ref{eq:G}) can be rewritten as
\begin{equation}
	\begin{split}
		x_{k+1} &= (A_G+\delta A_G)x_{k}+(B_G{}_1+\delta B_G{}_1) u_{k}\\
		&~~~~~~~~+(B_G{}_2+\delta B_G{}_2) \varphi_{k}\\
		&=A_Gx_{k} + B_G{}_1 u_k + B_G{}_2 \varphi_{k} + B_G{}_3 w_{k},
	\end{split}
\end{equation}
where $B_G{}_3 w:=\delta A_G x+\delta B_G{}_1 u+\delta B_G{}_2 \varphi$, with
\begin{align}
\begin{split}
&\delta A_G=
		\begin{bmatrix}
		  0 & 0 & 0 & 0 \\
		  0 & 0 & \epsilon \delta V_x & 0  \\
		  0 & 0 & 0 & 0 \\
		  0 & 0 & 0 & 0 \\
		\end{bmatrix},\\
		&\delta B_G{}_1 =
		\begin{bmatrix}
		  0 \\
		  -\epsilon\frac{l_r}{l_f + l_r} \delta V_x\\
		  0 \\
		  \frac{T}{(l_f + l_r)\tau_\psi}\delta V_x
		\end{bmatrix},
		\delta B_G{}_2 =
		\begin{bmatrix}
 		 0 & T\delta V_x\\ 0 & 0 \\ 0 & 0 \\ 0 & 0 \\
		\end{bmatrix}.
\end{split}
\end{align}
We suppose that there exists a scalar $\boldsymbol{\Delta}\in \mathcal{B}_{\boldsymbol{\Delta}}=\{ \boldsymbol{\Delta}  | \|\boldsymbol{\Delta}\|^2 \leq 1 \}$ and define the index $z$ as
\begin{align}
\begin{split}
v &:= C_G{}_1 x +D_G{}_{11} u+D_G{}_{12} \varphi,\\
w &:= \boldsymbol{\Delta} v.
\end{split}
\end{align}
The $w\rightarrow v$ channel is called the uncertainty channel.
We then see that  the following relationship such that
\begin{align}
\begin{split}
B_G{}_3w  &=
\delta A_G x+\delta B_G{}_1 u+\delta B_G{}_2 \varphi\\
&= \boldsymbol{\Delta} B_G{}_3  (C_G{}_1 x +D_G{}_{11} u+D_G{}_{12} \varphi)
\end{split}
\end{align}
is satisfied.
Therefore,  the state space fractional representation for the model uncertainty can be formulated as
\begin{align}
\begin{split}
\begin{bmatrix}
\delta A_G & \delta B_G{}_1  &\delta B_G{}_2
\end{bmatrix}=
\boldsymbol{\Delta} B_G{}_3
\begin{bmatrix}
C_G{}_1 & D_G{}_{11} & D_G{}_{12}
\end{bmatrix}.
\end{split}
\label{eq:LFT}
\end{align}
We assume that the bound of $\delta V_x$ can be found such that the following relationship is satisfied:
\begin{align}
	\norm{\delta V_x} \leq \overline{\delta V_x}.
\end{align}
Then the matrices of the fractional representation in (\ref{eq:LFT}) can be found
\begin{align}
\begin{split}
&B_G{}_3=
\begin{bmatrix}
1&0&0\\
0&1&0\\
0&0&0\\
0&0&1\\
\end{bmatrix},
C_G{}_1 =
\begin{bmatrix}
0 & 0 & 0 & 0\\
0 & 0 & \epsilon \overline{\delta V_x} & 0\\
0 & 0 & 0 & 0\\
\end{bmatrix},\\
&D_G{}_{11} =
\begin{bmatrix}
0 \\
-\epsilon \frac{l_r}{l_f+l_r} \overline{\delta V_x} \\
\frac{T}{(l_f+l_r)\tau_\psi} \overline{\delta V_x}
\end{bmatrix},
D_G{}_{12} =
\begin{bmatrix}
0 & T\overline{\delta V_x} \\
0 & 0\\
0 & 0
\end{bmatrix}.
\end{split}
\label{eq:define_LFT_matrix}
\end{align}
\subsection{Robust stability}
The uncertain plant $\bar{G}$ describing the vehicle lateral dynamics is defined as the following equations:
\begin{equation}
\begin{split}
	&\bar{G} =\\
	&\begin{cases}
            x_{k+1} &= A_Gx_{k} + B_G{}_1 u_k + B_G{}_2 \varphi_{k} + B_G{}_3 w_{k}\\
			v_k &= C_G{}_1 x_{k}+D_G{}_{11} u_k + D_G{}_{12} \varphi_k\\
		    y_k &= C_G{}_2 x_{k}
     \end{cases}\\
     &w_k = \boldsymbol{\Delta} v_k.
\end{split}
\label{eq:tilde_G}
\end{equation}

Defining
$s :=
\begin{bmatrix}
v^T
&
w^T
\end{bmatrix}^T
$, the vector $s$ then is represented as
\begin{align}
\begin{split}
s = D_{sv}v + D_{sw}w,
\end{split}
\label{eq:define_s}
\end{align}
where
$D_{sv} =
\begin{bmatrix}
I_{n_w} & 0
\end{bmatrix}^T
$, and
$D_{sw} =
\begin{bmatrix}
0 & I_{n_w}
\end{bmatrix}^T
$.
Defining an augmented state $\zeta := \begin{bmatrix}
		x^T& \xi^T
	\end{bmatrix}^T$, the dynamics of the augmented system is represented as
	\begin{equation}
		\begin{split}
			\zeta_{k+1} &= \mathcal{A} \zeta_k + \mathcal{B}_1 \varphi_k + \mathcal{B}_2 q_k + \mathcal{B}_3 w_k\\
			r_k &= \mathcal{C}_1 \zeta_k + \mathcal{D}_{12} q_k \\
			s_k &= \mathcal{C}_2 \zeta_k +
			\mathcal{D}_{21} \varphi_k +  \mathcal{D}_{22} q_k + \mathcal{D}_{23} w_k
		\end{split}\label{eq:uncertaincloseloop}
	\end{equation}
with
	\begin{equation*}
		\begin{split}
			&\mathcal{A} =
			\begin{bmatrix}
			A_G+B_G{}_1\tilde{D}_{12}C_G{}_2
				&
				B_G{}_1\tilde{C}_{1}\\
				\tilde{B}_{2}C_G{}_2
				&
				\tilde{A}
			\end{bmatrix},
			\mathcal{B}_1 =
			\begin{bmatrix}
			B_G{}_2\\
			0
			\end{bmatrix},\\
			&\mathcal{B}_2 =
			\begin{bmatrix}
			    B_G{}_1\tilde{D}_{11}\\
				\tilde{B}_1
			\end{bmatrix},
			\mathcal{B}_3 =
			\begin{bmatrix}
			    B_G{}_3\\
			    0
			\end{bmatrix},\\
			&\mathcal{C}_1=
			\begin{bmatrix}
				D_{rp}D_{22}C_G{}_2 & D_{rp}C_{2}
			\end{bmatrix},
			\mathcal{D}_{12}=D_{rq},\\
			&\mathcal{C}_2=
			\begin{bmatrix}
				D_{sv}(C_G{}_1+D_G{}_{11}\tilde{D}_{12}C_G{}_2)
				&
				D_{sv}D_G{}_{11}\tilde{C}_1
			\end{bmatrix},\\
			&\mathcal{D}_{21} =
			D_{sv}D_G{}_{12},
			\mathcal{D}_{22} =
			D_{sv}D_G{}_{11}\tilde{D}_{11},
			\mathcal{D}_{23} = D_{sw}.
		\end{split}
	\end{equation*}
Inspired by the work in~\cite{gu2022recurrent}, the next theorem merges the QC for $\tilde\Psi$ and LFT expression for $\Delta$ with the Lyapunov theorem to provide the exponential stability condition for the uncertain feedback system.
\begin{theorem}
Given a rate $\rho\in \mathbb{R}$ with $\rho\in(0,1]$, and $\Lambda \in \mathbb{D}^{n_\phi}_{+}$, the system (\ref{eq:uncertaincloseloop}) is $\rho$-exponentially stable, i.e., $\norm{x_k}\leq \sqrt{\text{cond}(P)}\rho^k \norm{x_0}$, $\forall k>0$, if there exist $P\in \mathbb{S}^{n_\zeta}_{++}$, $M_1\in \mathbb{S}^{2n_\Phi}_{++}$, $M_2\in \mathbb{S}^{2n_w}_{++}$, $\lambda_1>0$ and  $\lambda_2>0$ such that the following condition holds
	\begin{equation}
	\begin{split}
	&M_1
    \succ
    \lambda_1
    \begin{bmatrix}
    \Lambda   &   0  \\
    0   &    -\Lambda
    \end{bmatrix},
    M_2
    \succ
    \lambda_2
    \begin{bmatrix}
    I_{n_w}   &   0  \\
    0   &    -I_{n_w}
    \end{bmatrix},\\
		&\begin{bmatrix}
			I & 0 & 0 & 0\\
			\mathcal{A} & \mathcal{B}_1 & \mathcal{B}_2 & \mathcal{B}_3 \\
			\mathcal{C}_1 & 0 &  \mathcal{D}_{12} & 0\\
			\mathcal{C}_2 & \mathcal{D}_{21} &  \mathcal{D}_{22} & \mathcal{D}_{23}
		\end{bmatrix}^T
		\begin{bmatrix}
		    -\rho^2 P & 0 & 0 & 0 \\
		    0 & P & 0 & 0 \\
		    0 & 0 & M_1 & 0 \\
		    0 & 0 & 0 &  M_2
		\end{bmatrix}\\
		&~~~~~~~~~~~~~~~~~~~~~~~~~~~~~~~~\begin{bmatrix}
			I & 0 & 0 & 0\\
			\mathcal{A} & \mathcal{B}_1 & \mathcal{B}_2 & \mathcal{B}_3 \\
			\mathcal{C}_1 & 0 &  \mathcal{D}_{12} & 0\\
			\mathcal{C}_2 & \mathcal{D}_{21} &  \mathcal{D}_{22} & \mathcal{D}_{23}
		\end{bmatrix} \preceq 0.
		\label{eq:LMI1}
	\end{split}
	\end{equation}
\end{theorem}
~

\begin{proof}
Firstly, with (\ref{eq:QC}) and (\ref{eq:define_r}) we see that  the following relationship exists:
\begin{equation}
	r^T
	\begin{bmatrix}
		\Lambda & 0\\
		0 & -\Lambda
	\end{bmatrix}
	r \geq 0.
\end{equation}
Then, with some $\lambda_1 > 0$, it turns out that
\begin{align}
    r^T
	M_1
	r \geq 0
	\label{eq:M1_ineq}
\end{align}
is satisfied with
$M_1
    \succ
    \lambda_1
    \begin{bmatrix}
    \Lambda   &   0  \\
    0   &    -\Lambda
    \end{bmatrix}
$
using $\mathcal{S}$-procedure~\cite{boyd1994linear}. With $\boldsymbol{\Delta}\in\mathcal{B}_\Delta$ and (\ref{eq:define_s}), the following inequality exists:
\begin{equation}
	s^T
	\begin{bmatrix}
		I_{n_w}  & 0\\
		0 & -I_{n_w}
	\end{bmatrix}
	s \geq 0.
\end{equation}
Furthermore,  with some $\lambda_2 > 0$, we get that
\begin{align}
    s^T
	M_2
	s \geq 0
	\label{eq:M2_ineq}
\end{align}
is also satisfied with
$
M_2
    \succ
    \lambda_2
    \begin{bmatrix}
    I_{n_w}    &   0  \\
    0   &    -I_{n_w}
    \end{bmatrix}
$
using $\mathcal{S}$-procedure~\cite{boyd1994linear}. Multiplying on the left and right of (\ref{eq:LMI1}) by $\begin{bmatrix}
	    	\zeta^T_k & \varphi^T_k & q^T_k & w^T_k
	    \end{bmatrix}^T$ and its transpose leads to
	    \begin{equation}
	    	V(\zeta_{k+1})-\rho^2 V(\zeta_{k}) +
	    	r^T_k
	    	M_1
	    	r_k +
	    	s^T_k
	    	M_2
	    	s_k\leq 0.
	    	\label{eq:proof_ineq1}
	    \end{equation}
	    Let us define the Lyapunov function $V(\zeta):=\zeta^TP\zeta$.
	    With (\ref{eq:M1_ineq}) and (\ref{eq:M2_ineq}), iterating (\ref{eq:proof_ineq1}) to $k=0$ leads to
	    \begin{equation*}
	    		V(\zeta_{k}) \leq \rho^{2k}V(\zeta_{0}).
	    \end{equation*}
	    Therefore, we finally obtain
	    \begin{equation*}
	    	\norm{x_k} \leq \norm{\zeta_k}\leq \sqrt{\text{cond}(P)}\rho^k \norm{\zeta_0} \leq \sqrt{\text{cond}(P)}\rho^k \norm{x_0}.
	    \end{equation*}
\end{proof}
\begin{figure}[t]
    \centering
\includegraphics[width=0.5\columnwidth]{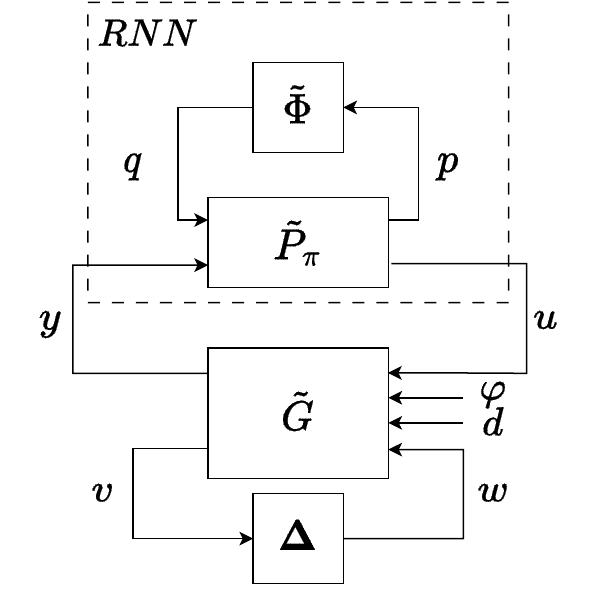}
		\caption{Feedback system consisting of uncertain plant $\tilde{G}$ and RNN controller.}
		\label{fig:fig3}
\end{figure}
\subsection{Reachable set analysis}
%
This subsection introduces the reachable set for the RNN-controlled closed-loop systems under model uncertainties and external perturbations, as shown in Fig.~\ref{fig:fig3}.
Introducing a perturbation $d$, the uncertain plant $\tilde{G}$ describing the vehicle lateral dynamics is then defined as the following equations:
\begin{equation}
\begin{split}
	&\tilde{G} =\\
	&\begin{cases}
            x_{k+1} &= A_Gx_{k} + B_G{}_1 u_k + B_G{}_2 \varphi_{k} + B_G{}_3 w_{k}+B_G{}_4 d_k\\
			v_k &= C_G{}_1 x_{k}+D_G{}_{11} u_k + D_G{}_{12} \varphi_k + D_G{}_{14} d_{k}\\
		    y_k &= C_G{}_2 x_{k}
     \end{cases}\\
     &w_k = \boldsymbol{\Delta} v_k.
\end{split}
\label{eq:tilde_G}
\end{equation}
\begin{assumption}
	The external disturbance $d$ is bounded such that $\norm{d}\leq d_{\text{max}}$.
	\label{assump:assum1}
\end{assumption}

With the augmented state
	$\zeta$, the dynamics of the augmented system is represented as
	\begin{equation}
		\begin{split}
			\zeta_{k+1} &= \mathcal{A} \zeta_k + \mathcal{B}_1 \varphi_k + \mathcal{B}_2 q_k + \mathcal{B}_3 w_k + \mathcal{B}_4 d_k\\
			r_k &= \mathcal{C}_1 \zeta_k + \mathcal{D}_{12} q_k \\
			s_k &= \mathcal{C}_2 \zeta_k +
			\mathcal{D}_{21} \varphi_k +  \mathcal{D}_{22} q_k + \mathcal{D}_{23} w_k + \mathcal{D}_{24} d_k
		\end{split}\label{eq:perturbedcloseloop}
	\end{equation}
with
	$
			\mathcal{B}_4 =
			\begin{bmatrix}
			    B_G{}_4\\
			    0
			\end{bmatrix}, \mathcal{D}_{24} = D_{sv}D_G{}_{14}.
	$

First, let us assume the stability condition (\ref{eq:LMI1}) for uncertainty is satisfied.
\begin{assumption}
	For a  $\rho\in \mathbb{R}$ with $\rho\in(0,1]$, there exist $P\in \mathbb{S}^{n_\zeta}_{++}$, $M_1\in \mathbb{S}^{2n_\Phi}_{++}$, $M_2\in \mathbb{S}^{2n_w}_{++}$, $\lambda_1>0$ and  $\lambda_2>0$ such that (\ref{eq:LMI1}) holds.
	\label{assump:assum2}
\end{assumption}
Then, for the perturbed uncertain feedback system (\ref{eq:tilde_G}), the next lemma provides an ellipsoidal outer approximation of the reachable set for the system state, similar to our previous method in~\cite{quan2023tube}.
\begin{proposition}
Consider the interconnection~(\ref{eq:tilde_G}) and let Assumption~\ref{assump:assum2} hold. There exists $\mu_d>0$ and $\mu_\varphi>0$ such that the following condition holds
	\begin{equation}
	\begin{split}
		&\Gamma^T
		\Pi\Gamma \preceq 0.
	\end{split}
	\label{eq:LMI2}
	\end{equation}
	with
	\begin{equation*}
		\begin{split}
		    &\Gamma = \begin{bmatrix}
			I & 0 & 0 & 0 & 0\\
			\mathcal{A} & \mathcal{B}_1 & \mathcal{B}_2 & \mathcal{B}_3 & \mathcal{B}_4\\
			\mathcal{C}_1 & 0 &  \mathcal{D}_{12} & 0 & 0\\
			\mathcal{C}_2 & \mathcal{D}_{21} &  \mathcal{D}_{22} & \mathcal{D}_{23}
			& \mathcal{D}_{24} \\
			0 & 0 & 0 & 0 & I_{n_d}\\
			0 & I_{n_\varphi} & 0 & 0 & 0
		\end{bmatrix},
		\end{split}
	\end{equation*}
	\begin{equation*}
		\begin{split}
			&\Pi = \begin{bmatrix}
		    -\rho^2 P & 0 & 0 & 0 & 0 & 0\\
		    0 & P & 0 & 0 & 0 & 0\\
		    0 & 0 & M_1 & 0 & 0 & 0\\
		    0 & 0 & 0 &  M_2 & 0 & 0\\
		    0 & 0 & 0 & 0 & -\mu_dI_{n_d} & 0 \\
		    0 & 0 & 0 & 0 & 0 & -\mu_\varphi I_{n_\varphi}
			\end{bmatrix}.
		\end{split}
	\end{equation*}
Further, for all $k\geq0$, it holds that
\begin{equation}
\zeta^T_k P \zeta_k
\leq \sigma_{k}
\end{equation}
where
\begin{align}
\begin{split}
\sigma_{0} &=
\zeta^T_0 P \zeta_0
\\
\sigma_{k+1} &= \rho^2 \sigma_{k} + \mu_d\norm{d_k}^2 +\mu_\varphi\norm{\varphi_k}^2
\end{split}
\label{eq:define_sigma}
\end{align}
\end{proposition}

~

\begin{proof}
Multiplying on the left and right by $\begin{bmatrix}
	    	\zeta^T_k & \varphi^T_k & q^T_k & w^T_k & d^T_k
	    \end{bmatrix}^T$ and its transpose leads to
	    \begin{equation}
	    \begin{split}
	    V(\zeta_{k+1})-\rho^2 V(\zeta_{k}) &+
	    	r^T_k M_1 r_k
	    	+
	    	s^T_k M_2 s_k \\
	    	&-\mu_d \norm{d_k}^2
	    	-\mu_\varphi \norm{\varphi_k}^2 \leq 0.
	   \end{split}
	   \label{eq:proof_ineq2}
	    \end{equation}
	    (\ref{eq:proof_ineq1}) guarantees that (\ref{eq:proof_ineq2}) holds with some $\mu_d>0$ and $\mu_\varphi>0$.
	    With (\ref{eq:M1_ineq}) and (\ref{eq:M2_ineq}) hold, the following inequalities are found
	   \begin{equation}
	   \begin{split} V(\zeta_{k+1})-\rho^2 V(\zeta_{k})
	    	-\mu_d \norm{d_k}^2
	    	-\mu_\varphi \norm{\varphi_k}^2\leq 0
	   \end{split}
	   \label{eq:V_ineq}
	   \end{equation}
	   By letting
$
	\sigma_{\kappa+1} = \rho^2 \sigma_{\kappa} + \mu_d\norm{d_\kappa}^2 + \mu_\varphi \norm{\varphi_\kappa}^2,
$
considering $\kappa \in [0, k-1]$,
the following inequalities can be obtained by multiplying (\ref{eq:V_ineq}) with $\rho^{2(k-\kappa-1)}$
\begin{align*}
\begin{split}
(V(\zeta_{k})-\sigma_{k})
-&\rho^2
(V(\zeta_{k-1})-\sigma_{k-1})
\leq 0,\\
\vdots\\
\rho^{2(k-1)}(V(\zeta_{1})-\sigma_{1})
-&\rho^{2(k-1)} \rho^2
(V(\zeta_{0})-\sigma_{0})
\leq 0.
\end{split}
\end{align*}
Summing up the inequalities above leads to
\begin{align}
\begin{split}
(V(\zeta_{k})-\sigma_{k})
-\rho^{2k}(V(\zeta_{0})-\sigma_{0})
\leq 0.
\end{split}
\end{align}
Then, we find that with (\ref{eq:define_sigma}), by choosing
$\sigma_{0} = \zeta^T_0 P \zeta_0$, there exists
\begin{equation}
	\zeta^T_k P \zeta_k = V(\zeta_{k})\leq\sigma_{k}.
\end{equation}
\end{proof}
For the lane-keeping tasks, we are interested in estimating the reachable set of the tracking error $e_{yL}$ so that the safety of the ego vehicle can be verified.
Here, we show how to find the upper bound for $e_{yL}$.
Let $P$ be decomposed into
$P=
\begin{bmatrix}
P_{11} & P^T_{21} \\
P_{21} & P_{22}
\end{bmatrix}
$ with
$
P_{11} \in \mathbb{R}^{1 \times 1}
$.
Defining
\begin{equation}
	P_{e_{yL}} = P_{11} - P^T_{21} P^{-1}_{22} P_{21},
\end{equation}
we can obtain
\begin{equation*}
	\begin{bmatrix}
		P_{e_{yL}} & 0\\
		0 & 0
	\end{bmatrix} =
	P -
	\begin{bmatrix}
		P^T_{21} P^{-1}_{22} P_{21} & P^T_{21} \\
		P_{21} & P_{22}
	\end{bmatrix}
	\preceq P,
\end{equation*}
with
$
\begin{bmatrix}
P^T_{21} P^{-1}_{22} P_{21} & P^T_{21} \\
P_{21} & P_{22}
\end{bmatrix}
\succeq 0
$, which can be derived by using the Schur complement for
$
P^T_{21} P^{-1}_{22} P_{21} - P^T_{21} P^{-1}_{22} P_{21} = 0 \succeq 0.
$
Finally, this yields
\begin{align}
e^T_{yL}{}_k P_{e_{yL}} e_{yL}{}_k
\leq
\zeta^T_k P \zeta_k
\leq
\sigma_{k}.
\end{align}

Noticing that using $\sigma_{k}$ defined in (\ref{eq:define_sigma}) requires the knowledge of the value of $d_{k}$ at each sample time, which is impractical in real-world applications.
Therefore, we introduce an upper bound $\bar \sigma_{k}$ of $\sigma_{k}$, in which $d_{\text{max}}$ is utilized instead of $d_{k}$.
The update law for $\bar\sigma_{k}$ is
\begin{align}
\bar\sigma_{k+1} = \rho^2 \bar\sigma_{k}
+\mu_d \norm{d_{\text{max}}}^2+\mu_\varphi \norm{\varphi_k}^2.
\end{align}

\begin{remark}
	Assumption~\ref{assump:assum2} guarantees the feasibility of the LMI condition (\ref{eq:LMI2}) due to the strict inequality, which provides infinite numbers of solutions.
	Suppose there exists $\gamma>0$ such that $P^{-1}_{e_{yL}}\leq \gamma$, which can be reformulated as the following LMI with Schur complement:
	\begin{equation}
		\begin{split}
		P^{-1}_{e_{yL}}\leq \gamma
		\Leftrightarrow
			\begin{bmatrix}
				P_{e_{yL}} & 1 \\
				1 & \gamma
			\end{bmatrix} \succeq 0
			\Leftrightarrow
			\begin{bmatrix}
				P_{22} & P_{21} & 0 \\
				P^T_{21} & P_{11} & 1\\
				0 & 1 & \gamma
			\end{bmatrix}\succeq 0.
		\end{split}
		\label{eq:LMI3}
	\end{equation}
	To find a minimum volume of the ellipsoidal reachable set, one can solve the following SDP problem:
	\begin{equation}
		\begin{split}
			&\min_{P,M_1,M_2,\lambda_1,\lambda_2} \gamma\\
			&~~~~~~~\text{s.t.}(\ref{eq:LMI2}),  (\ref{eq:LMI3})
		\end{split}
		\label{eq:SDP}
	\end{equation}
	with the values of $\mu_d$ and $\mu_\varphi$ fixed.
\end{remark}

\section{Experimental result}
\begin{figure}[t]
\centering
\SetFigLayout{2}{1}
  \subfigure[Training dataset]{\includegraphics[width=1\columnwidth]{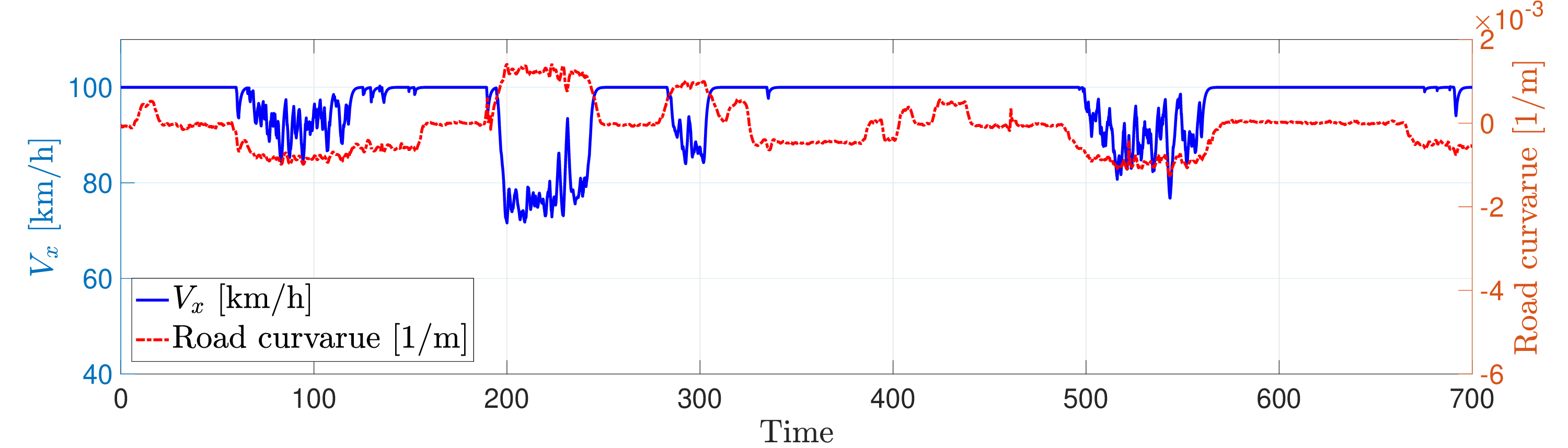}}
  \hfill
  \vspace{-0.1cm}
  \subfigure[Test scenario]{\includegraphics[width=1\columnwidth]{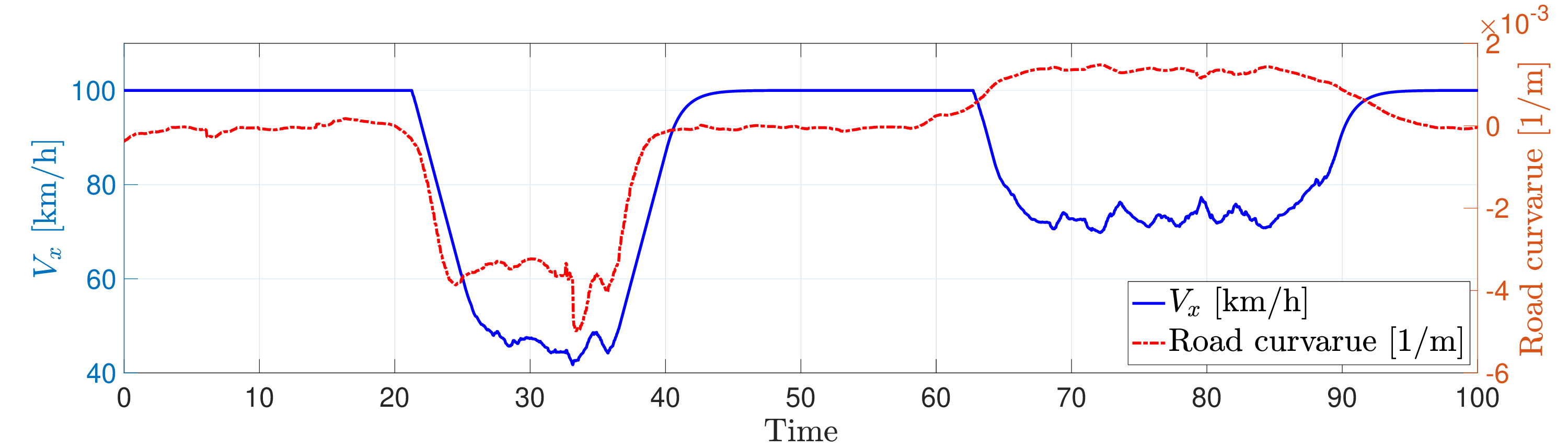}}
  \hfill
\caption{Vehicle velocity $V_x$ and road curvature.}
\vspace{-0.3cm}
\label{fig:road}
\end{figure}

\begin{figure}[t]
\centering
\SetFigLayout{2}{1}
  \subfigure[Training dataset]{\includegraphics[width=1\columnwidth]{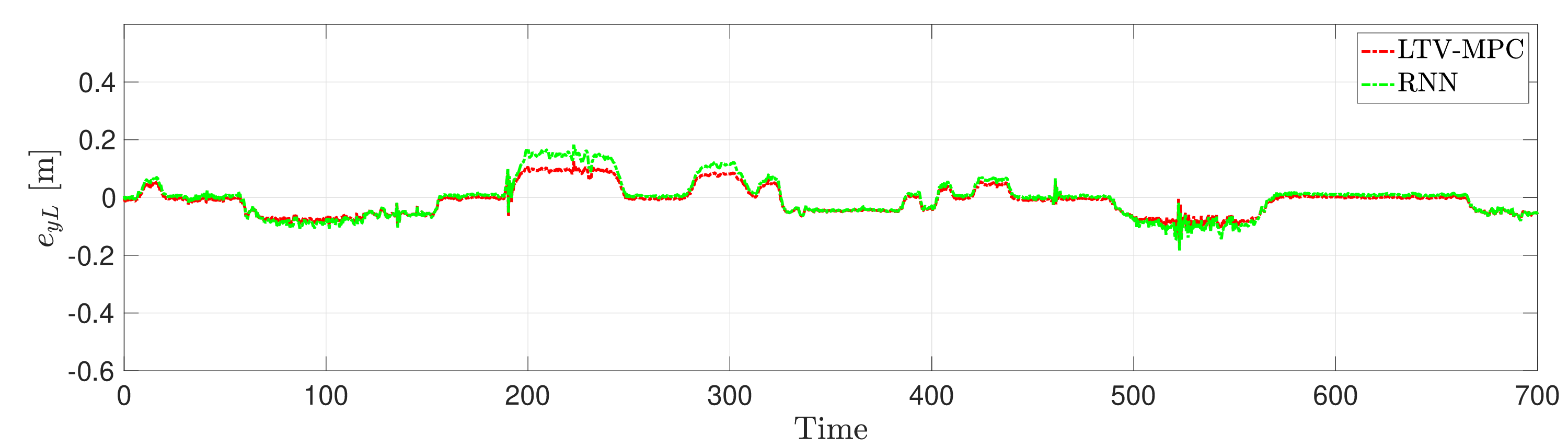}}
  \hfill
  \vspace{-0.1cm}
  \subfigure[Test scenario]{\includegraphics[width=1\columnwidth]{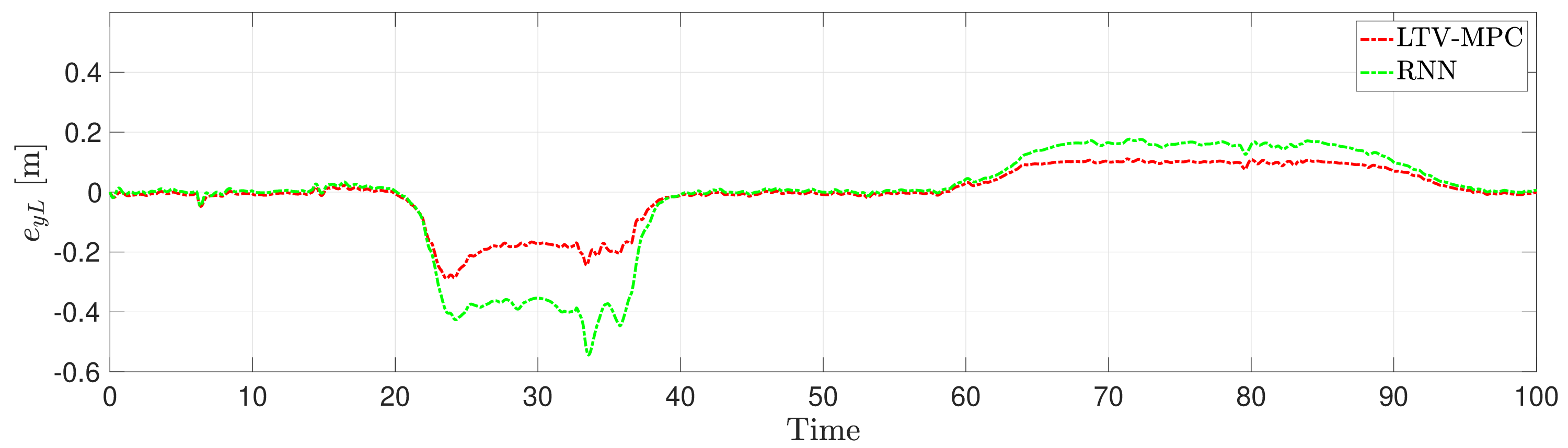}}
  \hfill
\caption{Tracking error $e_{yL}$ generated by LTV-MPC and RNN controller.}
\vspace{-0.3cm}
\label{fig:eyL}
\end{figure}

We validate our method in a lane-keeping scenario with varying vehicle speeds in MATLAB/Simulink.
The RNN controller was trained with Tanh activation functions in PyTorch~\cite{paszke2017automatic}.
The LMI in (\ref{eq:LMI2}) for stability verification and the SDP in (\ref{eq:SDP}) for reachable set calculation are solved using the YALMIP toolbox~\cite{lofberg2004yalmip}.
For the training data sets, an LTV-MPC is applied to the lane-keeping system, where the speed varies according to the road curvatures from \SI{70}{\kilo\metre\per\hour} to \SI{100}{\kilo\metre\per\hour} as shown in Fig.~\ref{fig:road} (a).
Road information for training datasets is collected by real vehicle experiments on national highway roads in Korea with both straight roads and roads with various curvatures concluded.
An LTV-MPC is designed assuming that the future speed variation in the control horizon is pre-obtained, which is infeasible in actual driving situations but possible in training phases where all the future information has already been included in the training datasets.
For the details of LTV-MPC design, readers are referred to our previous work~\cite{choi2021horizonwise}.
The test scenario introduces vehicle velocity and road curvatures not included in the training datasets, as shown in Fig.~\ref{fig:road} (b).
By fixing the decaying rate $\rho=0.9$, the robust stability of the RNN controller is verified by checking the feasibility of the LMI (\ref{eq:LMI1}).
From Fig.~\ref{fig:eyL}, we see that although the RNN controller exhibits similar performance with LTV-MPC with training dataset, the RNN controller generates a slightly different performance from LTV-MPC when encountering roads with curvatures not included in the training datasets.
However, with the LMI (\ref{eq:LMI1}) holds, we can verify that robust stability is maintained even with parameter uncertainty.
While LTV-MPC needs future speed information, which is impractical in real-world applications, on the other hand RNN needs only information from previous and current system states to give the output as the steering angle.

\begin{figure}[t]
    \centering
\includegraphics[width=1\columnwidth]{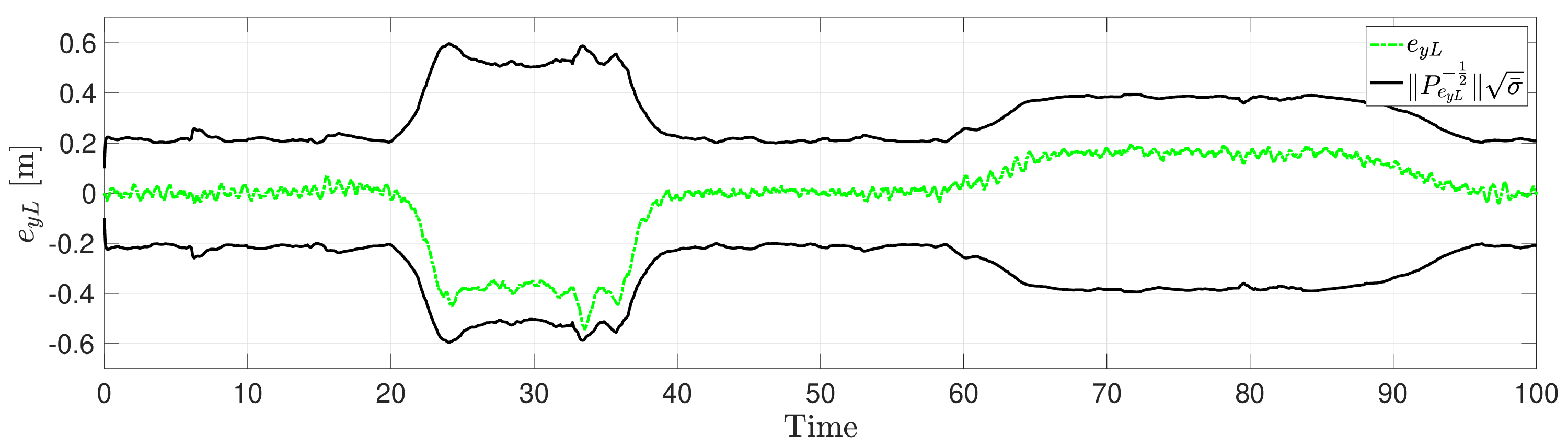}
		\caption{Tracking error $e_{yL}$ generated by RNN controller and its approximated bound in test scenario.}
		\label{fig:reach}
\end{figure}

With the LMI (\ref{eq:LMI1}) holds, the solutions for (\ref{eq:SDP}) are guaranteed.
We add external disturbance $d$ on the external input $\varphi$, satisfying $\norm{d} \leq \norm{d_{\text{max}}}$ with $d_{\text{max}} = \begin{bmatrix}
	0.01 & 0.01
    \end{bmatrix}^T$.
Fixing $\mu_d=\mu_\varphi=1$, the SDP (\ref{eq:SDP}) is solved.
As is shown in Fig.~\ref{fig:reach}, even in the test scenario with disturbances and untrained road conditions, we observed that the reachable set of $e_{yL}$ contains the actual $e_{yL}$, showing the effectiveness of the proposed method.
Therefore, with the proposed method, with untrained road conditions, the reachable set of system states can be validated and safety can be certified online.
Notice that the satisfaction of the LMI (\ref{eq:LMI1}) can be enforced via adding penalty terms in the objective function~\cite{revay2020convex,pauli2021training} or projection techniques~\cite{gu2022recurrent,junnarkar2022synthesis} during the learning process, which can be an improvement of this work.
For further details, readers are referred to~\cite{revay2020convex,gu2022recurrent,junnarkar2022synthesis,pauli2021training}.










%
\section{Conclusions}
This paper proposed an RNN controller for lane-keeping systems with dynamics uncertainty and external disturbance considered.
An LMI helps analyze the robust stability of the closed-loop system.
For untrained datasets with varying road curvatures, an outer approximation of the reachable set for the system state is provided for online safety verification.
Numerical experiments demonstrate the effectiveness of the proposed method.
For future work, training datasets with various road conditions will be employed to improve the performance of the proposed method.

\bibliographystyle{IEEEtran}


\end{document}